\documentclass[floatfix,aps,prl,groupedaddress,superscriptaddress,preprintnumbers,twocolumn]{revtex4}
\usepackage{graphicx}% Include figure files
\usepackage{bm}% bold math
\usepackage{amssymb,amsmath}

\begin{document}
	
%%%%%%%%%%%%%%%%%%%%%%%%%%%%%%%%%%%%%%%%%%%%%%%%%%%%%%%
	
	\title{Intrinsic multipole moments of the non-Gaussian wave packets}
	%\thanks{  }
	
	\author{Dmitry Karlovets} 
  %\email{d.karlovets@gmail.com}
\affiliation{Faculty of Physics, Tomsk State University, Lenina Ave. 36, 634050 Tomsk, Russia}
	
	\author{Alexey Zhevlakov} 
	%\email{zhevlakov@phys.tsu.ru}
	\affiliation{Faculty of Physics, Tomsk State University, Lenina Ave. 36, 634050 Tomsk, Russia} 
	\affiliation{Matrosov Institute for System Dynamics and Control Theory SB RAS Lermontov str. 134, 664033 Irkutsk, Russia} 

	\date{\today}
	
%%%%%%%%%%%%%%%%%%%%%%%%%%%%%%%%%%%%%%%%%%%%%%%%%%%%%%%	
	
\begin{abstract}
The charged wave packets with non-Gaussian spatial profiles are shown to possess intrinsic multipole moments.
The magnetic dipole moment and the electric quadrupole moment are found for a wide class of the packets, 
including the vortex electrons with orbital angular momentum $\ell$, the Airy beams, the so-called Schr\"odinger's cat states, and their generalizations.
For the packets with no phase vortices, the electric quadrupole moment is shown to grow quadratically with the packet's width, 
$|Q_{\alpha\beta}| \sim e\cdot \sigma_{\perp}^2$, while it is also $|\ell|$ times enhanced for the vortex beams.
For available beams of electron microscopes, these multipole moments are relatively easily adjusted and can be quite large, 
which affects the packets' electromagnetic fields and also allows one to develop new diagnostic tools 
for materials science, atomic and molecular physics, nuclear physics, and so forth.

\end{abstract}

% \pacs{41.75.Fr, 2.10.Dk, 31.30.jn, 21.10.Ky, 33.15.Kr}
% \keywords{wave packets, intrinsic moltipole moments }

\maketitle

\textit{Introduction.} -- An electron is known to have no electric dipole moment on a tree level because of the conservation of P- and T-invariance \cite{BLP}.
While its electric quadrupole moment vanishes too, it is so only as long as the wave packet itself is rotationally symmetric, and its transverse and longitudinal sizes coincide. 

In this Letter we point out that a non-vanishing magnetic dipole moment, an electric quadrupole moment, and the higher multipoles of a charged particle 
can also arise due to non-Gaussianity of the wave packet, that is, because of the specific quantum state either of a fermion or of a boson.
Vortex electrons carrying orbital angular momentum (OAM) $\ell$ with respect to the propagation axis \cite{Bliokh, Review}
and the Airy beams \cite{Airy, Airy_El_Exp} represent the simplest examples of such non-Gaussian states.
They have been recently generated at electron microscopes with the energies of 200-300 keV \cite{Airy_El_Exp, Uchida, Verbeeck, McMorran}, 
have found many applications in electron microscopy and tomography, in material studies, etc. \cite{Review},
and the vortex electron's OAM can reach the values of $\ell \sim 10^3 \hbar$ \cite{l1000}.
% Along with the electrons and photons, twisted neutrons have also been recently obtained \cite{neu}.

We calculate the first three intrinsic multipole moments of a charged packet as a function of its transverse profile or, in other words, 
of a wave function's phase $\varphi({\bm p})$. Based on a model of a so-called Bessel beam, it was previously pointed out by Bliokh \textit{et al.} 
that the vortex electrons possess a magnetic moment that is proportional to the OAM itself, thus becoming much larger than the Bohr magneton for high $\ell$ \cite{Bliokh}. 
The Bessel beam, however, is not a localized state, which is why it is unsuitable for estimating the higher multipole moments.
Here, we use a Laguerre-Gaussian (LG) beam instead, which represents an exact non-paraxial solution to the Schr\"odinger equation \cite{PRA} 
and allows one to calculate the quadrupole moment as well.

A wide class of the non-Gaussian packets is found to possess intrinsic magnetic dipole moments and electric quadrupole moments. The latter turn out to be 
an inherent property of the vortex packets, of the Airy beams, and of their different generalizations, including the so-called Schr\"odinger's cat states, 
which can be generated via the electron holography methods \cite{Review, Kruit}. These quadrupole moments are defined by the packet's spatial width $\sigma_{\perp}$, 
$Q_{\alpha\beta} \propto e\cdot\sigma_{\perp}^2$, and they can already be of the same order of magnitude as the corresponding moments of some atoms and molecules,
which opens up new venues for developing diagnostic tools, unavailable with the ordinary beams.
Thus, manipulating the spatial structure of the quantum waves allows one to produce particles carrying tailor-made multipole moments for the electron microscopy and tomography, 
atomic and molecular physics, material studies, and even hadronic and nuclear physics.

To separate the effects of non-Gaussianity from those arising due to the lack of rotational symmetry, we adhere to a model of a packet with equal transverse and longitudinal sizes. 
We also imply that the beam has a vanishing mean transverse momentum, $\langle{\bm p}\rangle = \{0,0,\langle p\rangle\}$. The system of units $\hbar = c = e = 1$ is used.

\textit{General definitions.} -- The first three moments of a system, ${\bm d}, {\bm \mu}$, and $Q_{\alpha\beta}$, are connected with a charge density $j^0$ and with a current density ${\bm j}$ as follows:
\begin{eqnarray}
& \displaystyle {\bm d} = \langle{\bm r}\rangle = \int d^3 r\, {\bm r}\, j^0 ({\bm r}),\ \ {\bm \mu} = \frac{1}{2}\int d^3r\, {\bm r}\times {\bm j} ({\bm r}),\cr
& \displaystyle Q_{\alpha\beta} = \int d^3r\, j^0 ({\bm r}) \left (3 r_{\alpha}r_{\beta} - {\bm r}^2 \delta_{\alpha\beta}\right).
\label{dmu}
\end{eqnarray} 
They can generally either be integrals of motion or depend on time. Staying within the non-relativistic framework, we ignore the possible dependence on the latter 
and, therefore, neglect the spreading. For a scalar packet with a mass $m$ and a wave function $\psi({\bm r}) = \int \frac{d^3p}{(2\pi)^3}\,\psi({\bm p})\, e^{i{\bm p}{\bm r}}$, the components of the current are
\begin{eqnarray}
& \displaystyle j^0 ({\bm r}) = |\psi ({\bm r})|^2,\ \int d^3r j^0 ({\bm r}) = 1,\cr 
& \displaystyle {\bm j} ({\bm r}) = \psi^*({\bm r})\frac{-i}{2m}{\bm\nabla}\psi({\bm r}) + \text{c.c.}
\label{j}
\end{eqnarray}

We start with a packet with a mean momentum $\langle{\bm p}\rangle$, a momentum uncertainty 
$$
\sigma_{p,x} = \sigma_{p,y} = \sigma_{p,z} = \sigma \equiv 1/\sigma_{\perp},
$$ 
the phase $\varphi ({\bm p})$, and the following wave function:
\begin{eqnarray}
& \displaystyle \psi ({\bm p}) = \Big ({\frac{2 \sqrt{\pi}}{\sigma}}\Big )^{3/2}\, \exp \Big \{-\frac{({\bm p} - \langle{\bm p}\rangle)^2}{2\sigma^2} + i \varphi ({\bm p})\Big\},\cr
& \displaystyle \int\frac{d^3p}{(2\pi)^3}\, |\psi ({\bm p})|^2 = \int d^3r j^0 ({\bm r}) = 1.  
\label{psi}
\end{eqnarray}
It is the phase $\varphi({\bm p})$ that makes the Fourier transform, $\psi({\bm r})$, non-Gaussian.

Substituting this into Eqs.(\ref{dmu}), (\ref{j}), one can easily derive how the multipole moments depend on the phase $\varphi$, 
\begin{eqnarray}
%& \displaystyle {\bm d}^{(beam)} (t) = N_b \left ({\bm r}_b + \langle {\bm u}\rangle t - \left\langle \frac{\partial \varphi ({\bm p})}{\partial {\bm p}}\right\rangle \right ) \equiv {\bm d}^{(beam)}_{ext}(t) + {\bm d}^{(beam)}_{int}, \cr
& \displaystyle {\bm d} = - \left\langle \frac{\partial \varphi ({\bm p})}{\partial {\bm p}}\right\rangle,
\
 {\bm \mu} = \frac{1}{2}\,\left\langle{\bm u}\times\frac{\partial \varphi ({\bm p})}{\partial {\bm p}}\right\rangle,
 \cr
& \displaystyle Q_{\alpha\beta} =  3 \left\langle \frac{\partial \varphi}{\partial p_{\alpha}}\frac{\partial \varphi}{\partial p_{\beta}}\right \rangle - \delta_{\alpha\beta} \left\langle \left (\frac{\partial \varphi}{\partial {\bm p}}\right )^2 \right \rangle,
\label{dnonrel}
\end{eqnarray}
where ${\bm u} = {\bm p}/m$ and the definition of a mean value of a test function $A ({\bm p})$ is
\begin{eqnarray}
& \displaystyle \langle A\rangle = \int \frac{d^3p}{(2\pi)^3}\, A({\bm p})\, |\psi ({\bm p})|^2.
\label{mean}
\end{eqnarray}
%If $A ({\bm p})$ is analytical in a vicinity of $\langle{\bm p}\rangle$, we have the following useful expansion in a series of $\sigma^2$:
%\begin{eqnarray}
%& \displaystyle \langle A\rangle = A\left (\langle{\bm p}\rangle\right ) + \frac{\sigma^2}{4}\, \frac{\partial^2 A({\bm p})}{\partial {\bm p}^2}\Big|_{{\bm p}=\langle{\bm p}\rangle} 
%+ \\
%& \displaystyle + \frac{\sigma^4}{32}\,\left (\left (\frac{\partial^2}{\partial {\bm p}^2}\right)^2 +2 \frac{\partial^4}{\partial {\bm p}^4}\right )A({\bm p})\Big|_{{\bm p}=\langle{\bm p}\rangle}  + \mathcal O (\sigma^6).
%\label{meanapp}
%\end{eqnarray}
%Clearly, the higher multipole moments depend on the higher derivatives of the phase. Hence, if $\varphi ({\bm p})$ is a polynomial, 
%the packet has only a finite number of the nonvanishing multipole moments.

Here we study the following three different types of the packets the first two of which have already been realized experimentally for electrons: 
\begin{itemize}
	 	\item
	 A packet with a singular phase -- the vortex beam with $\varphi_{\ell} = \ell \phi_p$ 
	 where $\phi_p$ is the momentum's azimuthal angle;
	\item
	 A packet with a non-singular phase -- the Airy beam with $\varphi_{\xi} = \frac{1}{3} \left(\xi_x^3 p_x^3 + \xi_y^3 p_y^3\right)$, 
	characterized by a vector ${\bm \xi } = \{\xi_x^3,\xi_y^3\}$;
	\item 
	 A quantum superposition of the Gaussian packets whose centers are spatially separated by a distance $2{\bm r}_0$ (the so-called Schr\"odinger's cat state).
\end{itemize}
% Possible means for the realization of the cat states are discussed in Ref.\cite{Kruit}.

%The derivatives of the corresponding phases are
%\begin{eqnarray}
%& \displaystyle \frac{\partial \varphi_{\xi}}{\partial {\bm p}} =\{\xi_x^3 p_x^2, \xi_y^3 p_y^2, 0\}\ (\text{Airy}),\\
%& \displaystyle \dfrac{\partial \varphi_{\ell}}{\partial {\bm p}} = \ell\, \dfrac{\hat{\bm z} \times {\bm p}}{{\bm p}_{\perp}^2}\ (\text{vortex}),\
%\hat{\bm z} = \{0,0,1\}.
%\label{dvarphi}
%\end{eqnarray}
%As can be seen, the derivative of $\varphi_{\ell}$ is not analytical at ${\bm p}_{\perp} \rightarrow 0$, even though the mean value (\ref{mean}) of $\partial \varphi_{\ell}/\partial {\bm p}$ is finite, 
%which leads to the finite dipole moments. However, for higher derivatives of $\varphi_{\ell}$ the singularity is no longer removable 
%and in this case one needs to amend the model (\ref{psi}) (see below). That is why the quadrupole moment of the vortex packet cannot be calculated within this approach.

\textit{Extrinsic vs. intrinsic contributions.} -- As the packet's charge is not vanishing, the dipole- and quadrupole moments embrace both intrinsic and extrinsic contributions.
While the former do not depend upon the choice of the coordinate origin, the latter moments are purely kinematic and can be eliminated merely by shifting the axes. 
To see this explicitly, let us shift the origin in the transverse plane by a vector ${\bm r}_0 = \{x_0, y_0, 0\}$: 
$$
{\bm r} \rightarrow {\bm r} + {\bm r}_0.
$$
As a result, the moments (\ref{dmu}) transform as follows: 
\begin{eqnarray}
& \displaystyle {\bm d} \rightarrow {\bm r}_0 + {\bm d}, \quad \ {\bm \mu} \rightarrow {\bm \mu} + \frac{1}{2}\int d^3r\, {\bm r}_0\times {\bm j} ({\bm r}),\cr
& \displaystyle Q_{\alpha\beta} \rightarrow Q_{\alpha\beta} + 3 (d_{\alpha}r_{0,\beta} + d_{\beta}r_{0,\alpha}) - \cr 
& \displaystyle - 2\,\delta_{\alpha\beta}\,{\bm d}{\bm r}_0 + 3\,r_{0,\alpha}r_{0,\beta}-\delta_{\alpha\beta}\,{\bm r}_0^2.
\label{dmushifted}
\end{eqnarray} 
%Although it seems generally impossible to find such a vector ${\bm r}_0$ that would turn the magnetic moment (i.e., the RHS in (\ref{dmushifted})) into zero,
%it vanishes when the current density can be (classically) written as ${\bm j} = j^0 \langle{\bm u}\rangle$ with $j^0$ sharply peaked at ${\bm r} = - {\bm r}_0$. 
%A similar situation takes place for the quadrupole moment: while a vector ${\bm r}_0$ that turns it into zero does not generally exist,
%it vanishes for some specific models of the packets (see below).

Let us now fix the origin,
$$
{\bm r}_0 = -{\bm d} = - \langle{\bm r}\rangle,
$$
so that in the new coordinates the packet rests at the origin on average, %\footnote{To be more precise, it moves with a constant speed, $\langle{\bm r}\rangle_{\text{int}} = \langle{\bm u}\rangle t$, 
%but the time dependence is irrelevant for the present discussion.}
$\langle{\bm r}\rangle_{\text{int}} = 0$.
It seems natural to call the multipole moments defined in this way \textit{the intrinsic moments} of the packet. 
They are
\begin{eqnarray}
& \displaystyle {\bm d}_{\text{int}} = 0, \quad \ {\bm \mu}_{\text{int}} = {\bm \mu} - \frac{1}{2}\int d^3r\, {\bm d} \times {\bm j} ({\bm r}),\cr
& \displaystyle Q_{\alpha\beta,\text{int}} = Q_{\alpha\beta} - 3 d_{\alpha}d_{\beta} + {\bm d}^2 \delta_{\alpha\beta},
\label{dmuQint}
\end{eqnarray} 
where ${\bm d}, {\bm \mu}$, and $Q_{\alpha\beta}$ are from (\ref{dmu}). In particular, for the packets from (\ref{psi}) we have\footnote{A proof that $\int d^3r {\bm j} = \langle{\bm u}\rangle$ is trivial only when the current can be represented as ${\bm j} = j^0 \langle{\bm u}\rangle$ and which is not the case for a vortex electron with $j_{\phi} \ne 0$. The latter component, however, does not contribute to the mean momentum, and the electron's motion stays rectilinear on average.}
%\begin{widetext}
\begin{eqnarray}
&& \displaystyle {\bm d}_{\text{int}} = 0,\ {\bm \mu}_{\text{int}} = \frac{1}{2}\,\left\langle{\bm u}\times\frac{\partial \varphi ({\bm p})}{\partial {\bm p}}\right\rangle - \cr
&& \displaystyle \qquad \qquad \qquad \quad - \frac{1}{2}\,\left\langle{\bm u}\right\rangle\times\left\langle\frac{\partial \varphi ({\bm p})}{\partial {\bm p}}\right\rangle,\cr
&& \displaystyle Q_{\alpha\beta,\text{int}} =  3 \left (\left\langle \frac{\partial \varphi}{\partial p_{\alpha}}\frac{\partial \varphi}{\partial p_{\beta}}\right \rangle - \left\langle \frac{\partial \varphi}{\partial p_{\alpha}}\right\rangle \left\langle \frac{\partial \varphi}{\partial p_{\beta}}\right\rangle\right ) - \cr
&& \displaystyle \qquad \qquad \quad - \delta_{\alpha\beta} \left (\left\langle \left (\frac{\partial \varphi}{\partial {\bm p}}\right )^2 \right \rangle - \left\langle \frac{\partial \varphi}{\partial {\bm p}}\right\rangle^2\right ).
\label{dmuQintmodel}
\end{eqnarray} 
%\end{widetext}
These moments are intrinsic by definition, and one can easily check this by shifting the coordinates' origin once again.
In this case, $\psi({\bm p}) \rightarrow \psi({\bm p}) \exp\{-i{\bm r}_0 {\bm p}\}$, 
and the phase $\varphi$ changes as
$$
\varphi \rightarrow -{\bm r}_0 {\bm p} + \varphi.
$$
Eq.(\ref{dmuQintmodel}) remains invariant under this transformation.

Thus, the non-Gaussian packets of charged particles do not have an intrinsic electric dipole moment, in accordance with the conservation of P- and T-invariance, 
but they can generally possess intrinsic magnetic dipole-, electric quadrupole-, and higher multipole moments.

\textit{Vortex packets.} -- For the vortex packet (\ref{psi}), we arrive at the following first two moments:
\begin{eqnarray}
& \displaystyle {\bm d}_{\text{int},\ell} = 0,\ {\bm \mu}_{\text{int},\ell} = \frac{\ell}{2m}\,\hat{{\bm z}},
\label{OAMrest}
\end{eqnarray}
in accordance with the result of Bliokh \textit{et al.} \cite{Bliokh, Review}. The moments in the laboratory frame of reference, 
in which the electron is relativistic with a mean energy $\langle\varepsilon\rangle = m/\sqrt{1-\beta^2}$, 
can be obtained from the Lorentz transformations for a 2nd rank tensor $({\bm d},{\bm \mu})$. %\footnote{To be more precise, this pair transforms as a product of an anti-symmetric tensor and a volume $V$. 
%This circumstance leads to some peculiarities for vortex particles beyond the paraxial approximation \cite{PRA}.}.
%\begin{eqnarray}
%\displaystyle {\bm d}_{\text{int},\ell} = \frac{\ell}{2m}\, {\bm \beta}\times\hat{\bm z},\
%{\bm \mu}_{\text{int},\ell} = \frac{\ell}{2m}\, \left (\hat{\bm z} - \frac{\gamma}{\gamma + 1}\, {\bm \beta} \beta_z\right ),
%\label{dmulab}
%\end{eqnarray}
In particular, for a packet with a non-vanishing transverse momentum there appears a finite electric dipole moment, 
${d}_{\text{int},\ell} = \beta_{\perp}\ell/(2m)$, while for a boost along the $z$ axis we have ${\bm d}_{\text{int},\ell} = 0,\, {\bm \mu}_{\text{int},\ell} = \hat{{\bm z}}\,\ell/(2\langle \varepsilon \rangle)$.

In order to calculate the quadrupole moment, we need to modify the model (\ref{psi}). 
As the azimuthal angle $\phi_p$ is not defined at $p_{\perp} = 0$, the Fourier transform of $\psi ({\bm p})$
decays non-exponentially at large distances, and the second moment diverges logarithmically. 
In order to restore exponential decay of the wave function with $\rho$, we add a prefactor of $p_{\perp}^{|\ell|}$ to $\psi({\bm p})$ 
and modify the normalization constant accordingly:
\begin{eqnarray}
& \displaystyle \psi_{\ell}({\bm p}) = \Big ({\frac{2\sqrt{\pi}}{\sigma}}\Big )^{3/2}\frac{p_{\perp}^{|\ell|}}{\sigma^{|\ell|}\sqrt{|\ell|!}}\exp \Big \{-\frac{({\bm p} - \langle{\bm p}\rangle)^2}{2\sigma^2} + \cr & \displaystyle \qquad + i\ell\phi_p\Big\}, \qquad \int\frac{d^3p}{(2\pi)^3}\,  |\psi_{\ell}({\bm p})|^2 = 1.  
\label{psiell}
\end{eqnarray}
% Note that this time $|\psi_{\ell}({\bm p})|^2$ depends on $|\ell|$, similarly to the Bessel beam \cite{Review}.
A Fourier transform of this state ($\bm{r} = \{\bm{\rho},z\}$),
\begin{eqnarray}
& \displaystyle \psi_{\ell}({\bm r}) = \int\frac{d^3p}{(2\pi)^3}\, \psi_{\ell}({\bm p})\, e^{i{\bm p}{\bm r}} = \frac{(i\rho)^{|\ell|}\sigma^{|\ell| + 3/2}}{\pi^{3/4}\sqrt{|\ell|!}}\cr
& \displaystyle \qquad \times\exp \Big \{i \langle p\rangle z + i\ell\phi_r  -\frac{\sigma^{2} r^2}{2}\Big\},
\label{Fourier}
\end{eqnarray}
represents a simplest LG beam with a radial index $n=0$ and taken at $t=0$ (cf.\,Eq.(5) in Ref.\,\cite{PRA}).
%Its density in the transverse plane represents a so-called \textit{gamma distribution} (see, for instance, \cite{Mandel}),
%\begin{eqnarray}
%& \displaystyle |\psi_{\ell}({\bm r})|^2 \propto \frac{(\rho\sigma)^{2|\ell|}}{|\ell|!}\,\exp\left\{-\sigma^2\rho^2\right\},  
%\label{gamma}
%\end{eqnarray}
%with its typical doughnut-like profile and a maximum at $\rho\sigma = \sqrt{|\ell|}$. 
If the time dependence is recovered, Eq.(\ref{Fourier}) obeys the Schr\"odinger equation \textit{exactly}, 
thus allowing one to quantitatively describe \textit{the non-paraxial effects}, 
in contrast to the relativistic LG states, which are only approximate (paraxial) solutions to the Klein-Gordon equation.

After these definitions, one can easily reproduce the dipole moments from Eq.(\ref{OAMrest}) and also derive the quadrupole moment of the vortex packet,
which is
\begin{eqnarray}
& \displaystyle Q_{\alpha\beta, \ell} = \langle\rho\rangle^2\, \text{diag} \{1/2, 1/2, -1\}, 
\label{Qell}
\end{eqnarray}
where
\begin{eqnarray}
& \displaystyle \langle\rho\rangle = \sqrt{|\ell|}\,\sigma_{\perp} \equiv \sqrt{|\ell|}/\sigma 
\label{Qell1}
\end{eqnarray}
is a mean radius of the packet \cite{PRA}. Here and in what follows we omit the subscript ``int''. 
This tensor is diagonal, traceless, it grows linearly with the OAM, and diverges in a plane-wave limit, $\sigma \rightarrow 0$. 
This divergence is a hallmark of the vortex beam. 
Thus, the quadrupole moment of the vortex electron,
\begin{eqnarray}
& \displaystyle
|Q_{\alpha\beta,\ell}| \sim e\cdot\langle\rho\rangle^2,
\label{Qellorder}
\end{eqnarray}
grows linearly with the OAM $\ell$, which is an effect of non-paraxiality.

%We would like to emphasize that this result is exact within the chosen model of the Laguerre-Gaussian beam (\ref{Fourier}) and in the non-relativistic approximation.
One can also try to reproduce this result with the unlocalized Bessel state for which $j^0 = N^2\,J_{\ell}^2(\kappa\rho)$ with $\kappa$ and $N$ being a transverse momentum and a normalization constant, respectively. Then we deal with an integral $I = \int_0^R d\rho\, \rho^3 J_{\ell}^2(\kappa\rho)$, 
which formally diverges as the normalization radius $R$ tends to infinity. Although one could guess from the dimension considerations that
$$
|Q_{\alpha\beta,\ell}| \propto N^2 I \propto \kappa^{-2},
$$
the integral over $z$ diverges too, thus making it hard to make quantitative estimates on the electron's quadrupole moment within the model of the Bessel beam.
To put it differently, the quadrupole- and higher multipole effects arise beyond the paraxial regime, for which only the well-localized LG beams (\ref{Fourier}) 
give reliable predictions.

\textit{Airy beams.} -- Unlike the vortex packet, the Airy beam has an azimuthally \textit{asymmetric} transverse profile \cite{Airy, Airy_El_Exp}.
However, after the averaging, the only non-vanishing intrinsic moment appears to be the quadrupole one, 
%\begin{widetext}
\begin{eqnarray}
& \displaystyle {\bm d}_{\xi} = {\bm \mu}_{\xi} = 0,\cr 
& \displaystyle Q_{\alpha\beta, \xi} = \frac{\sigma^4}{2}\,\text{diag} \{2\xi_x^6 - \xi_y^6,\, 2\xi_y^6 - \xi_x^6,\, -\xi_x^6 - \xi_y^6\},
\label{dmunonrelAiry}
\end{eqnarray}
%\end{widetext}
where we have returned to the model (\ref{psi}), applicable for a beam with a non-singular phase.
%Clearly, this tensor is also traceless and, unlike (\ref{Qell}), it vanishes for very wide beams, $\sigma_{\perp}\rightarrow \infty$.
Although this moment is attenuated as $\sigma^4$ and seems to vanish in the plane-wave limit, $\sigma \rightarrow 0$, 
the maximum value of $\xi_x,\xi_y$ is smaller than the packet's width $\sigma_{\perp} \equiv 1/\sigma$ \cite{JHEP}. 
As a result,
\begin{eqnarray}
& \displaystyle |Q_{\alpha\beta,\xi}| \sim e\cdot\xi_{x,y}^6\sigma^4\lesssim e\cdot\sigma_{\perp}^2,
\label{QAiryestim}
\end{eqnarray}
which is roughly $|\ell|$ times weaker than the corresponding moment (\ref{Qellorder}) of the vortex electron.

\textit{Schr\"odinger's cat state.} -- Let us take a quantum superposition of two Gaussian packets without phases whose centers are shifted in the transverse plane by a distance ${\bm r}_0 = \{x_0, y_0, 0\}$.
Such states of electrons can come in handy for the so-called interaction-free measurements \cite{Elitzur} and they can be generated in near future % at the electron microscopes 
\cite{Kruit}. The normalized wave function of this packet, dubbed the Schr\"odinger's cat state, is
\begin{eqnarray}
& \displaystyle \psi_{1\pm 1} ({\bm p}) = \frac{1}{\sqrt{2}}\frac{e^{-i{\bm r}_0{\bm p}} \pm e^{i{\bm r}_0{\bm p}}}{\sqrt{1 \pm \exp\{-\sigma^2 r_0^2\}}}\, \psi({\bm p}), 
%\cr & \displaystyle \int\frac{d^3p}{(2\pi)^3}\, |\psi_{1\pm 1} ({\bm p})|^2 = 1,
\label{catpsi}
\end{eqnarray}
where $\psi({\bm p})$ is from Eq.(\ref{psi}) with $\varphi = 0$. 
The upper sign corresponds to the so-called \textit{even} cat state and the lower one to the \textit{odd} state.
Evaluating the current in Eq.(\ref{j}), we arrive at the following result:
\begin{eqnarray}
& \displaystyle {\bm d}_{1\pm 1} = {\bm \mu}_{1\pm 1} = 0,\\ 
& \displaystyle Q_{\alpha\beta, 1\pm 1} = \frac{1}{1\pm\exp\{-\sigma^2r_0^2\}} \left(3r_{0,\alpha}r_{0,\beta} - {\bm r}_0^2 \delta_{\alpha\beta}\right).
\label{dmucat}
\end{eqnarray}

Thus, the cat state, too, has only the quadrupole moment, which is proportional to the distance between the packets, $r_0 \gtrsim \sigma_{\perp}$.
As this distance ${\bm r}_0$ itself does not depend on the choice of the origin, this moment is intrinsic. 
A rough estimate of it is
\begin{eqnarray}
& \displaystyle |Q_{\alpha\beta, 1\pm 1}| \sim e\cdot r_0^2 \gtrsim e\cdot \sigma_{\perp}^2 \sim |Q_{\alpha\beta,\xi}|,
\label{dmucatrough}
\end{eqnarray}
which is of the same order of magnitude as the corresponding moment of the Airy beam (\ref{QAiryestim}).

\textit{Discussion.} -- As can be seen from Eq.(\ref{dmuQintmodel}), the magnetic moments are generally non-vanishing for non-Gaussian packets. 
The vortex beam, therefore, is not the only state that reveals this property -- one can generate other beams with the more sophisticated profiles
and magnetic moments by using the electron holographic technologies. Different kinds of the non-Gaussian beams are discussed nowadays in optics \cite{AB_2010}. 
Moreover, it turns out that possession of an electric quadrupole moment 
\textit{is rather a rule than an exception} for such a packet. For Airy packets and other beams with the non-singular phases, 
these moments are $|Q_{\alpha\beta}| \sim e\cdot \sigma_{\perp}^2$. 
The widths of the packets of electron microscopes can span from the subnanometer range \cite{Angstrom} to tens of microns,
which yields
$$
|Q_{\alpha\beta}| \sim 10^{-16} - 10^{-6}\, e\cdot \text{cm}^2,
$$
and $|\ell|$ times higher for the vortex beams. In particular, the packet focused to a spot of a Bohr radius $a_B \sim 0.1$ nm in diameter \cite{Angstrom} 
has a moment $e\cdot a_B^2 \sim 10^{-16}\, e\cdot \text{cm}^2$, which is of the same order of magnitude as the quadrupole moments of light hydrogen-like atoms 
in a $2P_{3/2}$ state \cite{Qatom} or of some diatomic molecules \cite{Qmol}. Thus, such non-Gaussian packets can be used for probing 
the quadrupole moments of atomic and molecular systems, as well as their quadrupole polarizability, due to the quadrupole-quadrupole interaction. 
Its intensity can be enhanced by controlling the beam's width and the parameters.

Furthermore, the beams' quadrupole moments can be more sensitive to the quadrupole polarizability of the materials with spatial dispersion, 
thus allowing one to go beyond a dipole approximation in scattering of the electron waves by surfaces and artificial structures, 
including recently created metamaterials with a bulk quadrupole moment \cite{Garcia}. The usually weak quadrupole transitions in atoms are known to be enhanced when the latter interact with the twisted photons \cite{Babiker}, and we envisage similar enhancement of the quadrupole effects in both elastic and inelastic scattering of the non-Gaussian sub-nanometer-sized electron waves 
by atoms and molecules, as a parameter governing the quadrupole phenomena, $(a_B/\sigma_{\perp})^2$, ceases to be small.

The beams endowed with the multipole moments can also find applications in electron tomography and holography for materials science. 
Indeed, it is the OAM-induced magnetic moment of the vortex beam that has recently allowed for the measurements of all three components of a nanoscale magnetic field
without tilting a specimen \cite{Grillo}. While the magnetic moment couples to the magnetic field inside a specimen as ${\bm \mu}\cdot{\bm H}$ and 
the electric dipole moment couples to the electric field, ${\bm d}\cdot{\bm E}$, the packet with a quadrupole moment allows one to probe \textit{the field gradients} inside a material,
as the potential energy is $-\frac{1}{6}\,Q_{\alpha\beta}\partial E_{\alpha}/\partial x_{\beta}$. %\cite{L2}. 
Therefore, the non-Gaussian packets can also serve as a means for a 3D imaging of the fields inside a specimen.
Then, by employing coherent superpositions of such states, this imaging can be made non-invasive, 
thus realizing an idea of the so-called interaction-free measurements \cite{Elitzur, Kruit}, which can be of crucial importance for studying biological systems.

For the moment, twisted neutrons are the only hadrons with a non-Gaussian profile that have been obtained \cite{Clark}.
However, if the charged vortex- or Airy hadrons -- say, protons -- were created, they would carry quadrupole moments several orders of magnitude lower than electrons,
and this would allow one to use them for probing quadrupole moments of the nuclei, which do not exceed a few barns. 
A simplest way to obtain a vortex proton is to get the OAM transferred from the vortex electron
scattered by the proton elastically at a small angle, $e_{\ell} + p \rightarrow e + p_{\ell}$. 
If the final electron is projected on a plane-wave state, the proton will carry the OAM just due to the conservation law. 
The moderately relativistic electrons with the energies of several MeV would suffice for these purposes.
% On the contrary, the ultrarelativistic vortex- or Airy electrons with the energies of several hundreds of MeV can become useful in measurements 
% of the charge radii and the quadrupole moments of nuclei by means of electron-nucleus scattering.

\begin{figure}
\centering
\includegraphics[width=7.50cm, height=4.40cm]{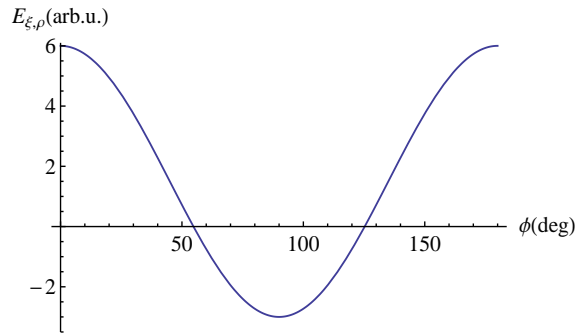}
\caption{Unlike the vortex packet, the Airy beam lacks azimuthal symmetry. This results in an azimuthal dependence 
of the radial field $E_{\xi,\rho}$ (\ref{EQAiry}) of its quadrupole moment (\ref{dmunonrelAiry}). Parameters are $\eta = 0\, (\xi_y = 0), \theta = \pi/2$.}
\label{Asymm}
\end{figure}

The packet's multipole moments also produce electromagnetic fields. At large distances, where the higher multipoles can be neglected, 
its electric field represents a sum ${\bm E} = {\bm E}_e + {\bm E}_Q$, whereas the magnetic field is just that of the magnetic dipole, ${\bm H} = {\bm H}_{\mu}$, 
which for a dipole at rest is ${\bm H}_{\mu} = (3{\bm n} ({\bm n}{\bm \mu}) - {\bm \mu})/r^3,\, {\bm n} = {\bm r}/r = \{\sin\theta\cos\phi,\sin\theta\sin\phi,\cos\theta\}$. 
The field of the quadrupole,
\begin{eqnarray}
& \displaystyle
 E_{Q,\alpha} = \frac{5}{2}\, r_{\alpha} \frac{r_{\beta} r_{\gamma} Q_{\beta\gamma}}{r^7} - \frac{Q_{\alpha\beta}r_{\beta}}{r^5},\ \alpha = 1,2,3,
\label{EQ}
\end{eqnarray}
stays azimuthally symmetric for the vortex packet, ${\bm E}_{\ell} = E_{\ell,\rho}\, \hat{\bm \rho} + E_{\ell,z}\, \hat{\bm z},\, E_{\ell,\phi} = 0$, similarly to the Bessel beam \cite{Lloyd}. 
Its nonvanishing components are %\footnote{Note that the fields of the relativistic packet in the laboratory frame can be easily obtained by the Lorentz transformations.} 
\begin{eqnarray}
& \displaystyle E_{\ell,\rho} = \frac{3}{4}\frac{\langle\rho\rangle^2}{r^4} \sin\theta\, (1 - 5 \cos^2\theta),\cr
& \displaystyle E_{\ell,z} = \frac{3}{4}\frac{\langle\rho\rangle^2}{r^4} \cos\theta\, (3 - 5 \cos^2\theta). 
\label{EQOAM}
\end{eqnarray}
The magnetic field ${\bm H}_{\ell} = {\bm H}_{\mu} = H_{\ell,\rho}\, \hat{\bm \rho} + H_{\ell,z}\, \hat{\bm z}$, unlike that of the Bessel beam \cite{Lloyd}, has a non-vanishing component $H_{\ell,\rho}$. 
Importantly, these fields \textit{grow linearly with the OAM}, as do both $\langle\rho\rangle^2 = |\ell|\, \sigma_{\perp}^2$ and $\mu_{\ell} = \ell/2m$.
This enhancement represents yet another non-paraxial effect, which does not take place for the Bessel beam \footnote{The Bessel beam, being delocalized both spatially and temporarily, 
does not in fact allow one to adequately describe electromagnetic fields of a vortex electron. 
Although the solution of Ref.\cite{Lloyd} seems to be formally correct, its magnetic field has no $H_{\rho}$-component and does not coincide with that of a magnetic dipole at large distances.
The physically consistent fields can be found by making use of the well-localized LG states.}. 
 
% $E_{\rho} = E_{e, \rho} + E_{\ell,\rho} = r^{-2}\sin\theta\,(1 + \frac{3}{4}\frac{\langle\rho\rangle^2}{r^2}(1-5\cos^2\theta))$.

On the contrary, the quadrupole field of a packet with no azimuthal symmetry has all three components.
For the Airy beam with ${\bm \xi} = \xi^3 \{\cos\eta,\sin\eta\}$ we have instead
%\begin{widetext}
\begin{eqnarray}
% & \displaystyle {\bm E}_{\xi} = E_{\xi,\rho}\, \hat{\bm \rho} + E_{\xi,\phi}\, \hat{\bm \phi} + E_{\xi,z}\, \hat{\bm z},\cr
&& \displaystyle E_{\xi,\rho} = \frac{1}{4}\frac{\sigma^4 \xi^6}{r^4} \sin\theta\, (3 - 5 \cos^2\theta) \Big(2 - 3\cos^2\eta + \cr
&& \displaystyle \qquad \qquad \qquad + 3\cos^2\phi(2\cos^2\eta-1)\Big),\cr
&& \displaystyle E_{\xi,z} = \frac{1}{4}\frac{\sigma^4 \xi^6}{r^4} \cos\theta \Big(5\, (3\sin^2\theta\cos^2\eta\cos(2\phi) + \cr
&& \displaystyle \qquad \qquad + \sin^2\theta (2 - 3\cos^2\phi) - \cos^2\theta) + 2\Big),\cr
&& \displaystyle E_{\xi,\phi} = \frac{3}{2}\frac{\sigma^4 \xi^6}{r^4} \sin \theta \cos\phi \sin\phi \cos(2\eta), 
\label{EQAiry}
\end{eqnarray}
%\end{widetext}
where both $E_{\xi,\phi}$ and the dependence on $\phi$ of the other components vanish when $\eta = \pi/4$. 
In Fig.\ref{Asymm} we show the azimuthal asymmetry of the field's radial component, which is quite big and can be reliably detected.

Thus, by measuring the fields of a particle in a special quantum state one can in principle retrieve its multipole moments
and, thereby, perform tomography of the quantum state itself.
%find the first several moments of the packet's phase. % Subsequently, the phase $\varphi({\bm r})$ of the wave in x space can also be found, up to a constant term.
Such a scheme can be used in addition to a holographic method, recently employed for the retrieval of the phase of the vortex beam in Ref.\cite{Venturi}.

% Finally, we note that within the ring, $r\lesssim\langle\rho\rangle$, both the quadrupole's field strength and its gradient exceed those of the bare charge.

% As a result, finite electric quadrupole moments of the charged particles, if found, may not necessarily have arisen thanks to the new physics but merely due to specific quantum state of the wave packet itself,
% even though it seems highly unlikely that such a state can spontaneously arise in some physical process. On the other hand, it is known that at least the vortex states of light (and certainly of matter) can be  naturally generated, for instance, in a vicinity of the rotating black holes \cite{Tamburini}. One cannot exclude, therefore, that there are other processes in which the packets become non-Gaussian and acquire the finite electric moments -- say, via diffraction and scattering on the inhomogeneous periodic structures and surfaces. $\leftarrow$ \textit{Any ideas in astrophysics!!!???}

To conclude, we have shown that the non-Gaussian charged packets generally possess multipole moments, 
if they are not excluded by the invariance considerations. In particular, the electric quadrupole moments of the vortex beams, the Airy packets and of their generalizations 
have been shown to be quite large, which makes them potentially useful for the development of the new diagnostic tools, unavailable with the ordinary beams.
This includes diagnostics of materials, atoms and molecules, nuclei, and, by measuring electromagnetic fields of such beams, tomography of the quantum state of the particles. 

We are grateful to V.~Bagrov, I.~Ivanov, P.~Kazinski, V.~Serbo, A.~Shishmarev, and A.~Tishchenko for many fruitful discussions and criticism. 
This work is supported by the Russian Science Foundation (project No.\,17-72-20013).

\end{document}